\documentclass[12pt, anon]{l4dc2023}
\usepackage[utf8]{inputenc}
\usepackage{mathtools}
\usepackage{amsmath,amssymb,amsfonts}
\usepackage{amsmath,amssymb,amsfonts}
\usepackage{mathrsfs}
\usepackage{graphicx}
\usepackage{textcomp}
\usepackage{xcolor}
\usepackage{algorithm}
\usepackage{algpseudocode}
\usepackage{comment}
\usepackage{ragged2e}

\usepackage{wrapfig, blindtext}

\newtheorem{assumption}{Assumption}

\title[Time Dependent IOC]{Time Dependent Inverse Optimal Control \\ using Trigonometric Basis Functions}
\usepackage{times}

\author{\Name{Rahel Rickenbach} \Email{rrahel@ethz.ch}\\
  \Name{Elena Arcari} \Email{earcari@ethz.ch}\\
  \Name{Melanie N. Zeilinger} \Email{mzeilinger@ethz.ch}\\
  \addr ETH Zurich
}

\begin{document}

\maketitle

\begin{abstract}%
The choice of objective is critical for the performance of an optimal controller. When control requirements vary during operation, e.g. due to changes in the environment with which the system is interacting, these variations should be reflected in the cost function. 
In this paper we consider the problem of identifying a time dependent cost function from given trajectories. We propose a strategy for explicitly representing time dependency in the cost function, i.e. decomposing it into the product of an unknown time dependent parameter vector and a known state and input dependent vector, modelling the former via a linear combination of trigonometric basis functions. These are incorporated within an inverse optimal control framework that uses the Karush–Kuhn–Tucker (KKT) conditions for ensuring optimality, and allows for formulating an optimization problem with respect to a finite set of basis function hyperparameters. Results are shown for two systems in simulation and evaluated against state-of-the-art approaches.\footnote{The datasets generated and/or analysed during the current study are available in the eth research collection repository, https://doi.org/10.3929/ethz-b-000611670.}
\end{abstract}


\section{Introduction}
\label{sec:introduction}
The performance of an optimization-based controller significantly depends on the ability to encode the desired goal in the cost function. This can be challenging in several scenarios, e.g. in the context of autonomous driving~\cite{MK15}, or for biomedical applications~\cite{khodaei2020}. Particularly for the latter, the ability to specify changes in the control requirements that reflect, e.g., variations of the human body in time, is an essential condition for successfully fulfilling the task at hand.
For instance, maintaining the basal insulin requirement is a concrete example that
shows both increased variability during day and night time, and time-varying profiles over the day~\cite{ruan2016,scheiner2005}. Among the non-biomedical applications, another example is given by the optimal control with respect to continuously changing electricity prices~\cite{Mbung2017}. 
Motivated by these examples, in this paper we tackle the problem of specifying a cost function that explicitly depends on time. To this end, we build on the inverse optimal control (IOC) approach presented in~\cite{menner2019}, which encodes optimality conditions by exploiting the Karush–Kuhn–Tucker (KKT) conditions~\cite{kuhn2014} and Bellman’s Principle of Optimality \cite{bellman1966}. Within this framework, we incorporate the learning of \textit{time features} that we model as trigonometric basis functions, taking inspiration from~\cite{lazaro2010} and~\cite{arcari2021}: their inherent nonlinearity provides a sufficiently rich description of the objective's time dependency, while preserving the efficiency of parametric models. \\ The  objective is assumed to decompose as the product of an unknown continuous-time parameter vector, and a known state and input dependent vector of arbitrary structure. The unknown time vector is defined as a linear combination of trigonometric basis functions, which allows for formulating the time feature extraction as an optimization problem with respect to a finite set of hyperparameters, e.g. sinusoidal frequencies. These are jointly optimized with the Lagrangian variables arising in the IOC problem formulation. Consequently, the proposed algorithm solves the Lagrangian optimization problem regarding the unknown cost function parameters, as well as a line search over a regularization term for obtaining a sparse solution. Due to non-convexity of the optimization problem, a grid search over initial conditions is performed. Results are shown for two simulation examples, i.e. a multi-layer spring-damper system, and an inverted double pendulum, for which performance is evaluated against state-of-the-art approaches. Furthermore, the reliability of our estimates is investigated by varying both sampling times and horizon lengths of the forward control problem. \\ The preliminaries are summarized in Section~\ref{sec:preliminaries}, and the problem description is given in Section~\ref{sec:problemdescription}. This is followed by an explanation of the proposed approach in Section~\ref{sec:tblrbasedtimedependentioc}; while the first subsection is devoted to the introduction of the considered learning approach, the final algorithm follows in Section~\ref{sec:featuredependentioc}. Results are gathered in Section~\ref{sec:results}, and we conclude the paper with a discussion in Section~\ref{sec:conclusion}.

\subsection{Related Work}
\label{sec:relatedwork}
The use of optimal demonstrations to learn the parameters of an initially unknown objective has been widely addressed in the literature \cite{FengShunLin:2021}, both from a reinforcement learning and an optimal control perspective~\cite{abazar2020}. Inverse reinforcement learning (IRL) is often formulated for discrete state and action spaces, and mainly focuses on the reconstruction of either a reward function or a policy from expert demonstrations, and therefore is also considered in the framework of \textit{learning from demonstrations} or \textit{imitation learning}~\cite{ng2000},~\cite{ramachandran2007},~\cite{arora2021}. On the other hand, IOC exploits the structure of the forward optimization problem to formulate the corresponding cost function learning approach by typically using stability and/or optimality conditions. Regarding the latter, a large variety of objective estimation techniques were developed: for example, the works of~\cite{priess2014} and~\cite{menner2018} are inspired by the solution of an LQR problem, while~\cite{li2011} relies on the Hamiltonian Jacobi Bellman equation to estimate the unknown objective. Overall, these approaches focus on infinite-horizon optimization problems without considering the presence of potential constraints. An IOC approach including inequality constraints was presented in~\cite{englert2017} by exploiting KKT optimality conditions. Building on this idea, the method presented in~\cite{menner2019} additionally exploits Bellman's Principle of Optimality in order to solve the original infinite-horizon formulation using finite-length demonstration trajectories. Other extensions focus on the consideration of uncertain data and noise, e.g. in \cite{menner2020}, and also on the estimation of time-varying objectives \cite{jin2019}, \cite{westermann2020}, \cite{lin2016}.  
While in these works time dependency is integrated by either separating the considered time horizon into windows with constant parameters, or by averaging over multiple windows, the method presented in this paper explicitly models time as a cost function feature, while preserving the consideration of constraints. 
The approach of estimating individual cost parameters at each time step, as well as the non-parametric KKT approach, both presented in~\cite{englert2017} also allow for learning time-varying cost functions in the finite-horizon setting. 
However, time dependency is incorporated via a time-varying vector of parameters, whose length matches the duration of the task at hand, and can therefore potentially become prohibitively long. In this work, the use of time features bypasses this issue since it directly provides a \textit{model} for the relation between cost function and time to be used in the infinite-horizon setting.

\subsection{Notation}
\label{sec:notation}
Throughout the paper $\Vert \cdot \Vert$ indicates the Euclidean norm, while $\vert \cdot \vert$ indicates the 1-norm. When applied to a matrix, the latter refers to the sum over all absolute values of its elements. The set of all non-negative real numbers is indicated with $\mathbb{R}_{+}$.

\section{Preliminaries}
\label{sec:preliminaries}
In this work, we build upon the \textit{shortest path} IOC (spIOC) approach developed by \cite{menner2019}, considering $D>0$ finite-length observations provided by a demonstrator. These are assumed to be optimal trajectory segments of the original infinite-horizon constrained optimal control problem, and consist of state and input measurements at time instances $k \in \mathbb{N}$, with $x_{d}^{*}(k) \in \mathbb{R}^{n}$ and $u_{d}^{*}(k) \in \mathbb{R}^{m}$, $d\in[1,D]$, collected for different initial conditions $x_{d}^{*}(k)$ over a horizon $N \in \mathbb{N}$. The resulting sequences, indicated by $\mathcal{X}_{d}^{*} = \{x_{d}^{*}(k), \hdots, x_{d}^{*}(k+N)\}$ and $\mathcal{U}_{d}^{*} = \{u_{d}^{*}(k), \hdots, u_{d}^{*}(k+N-1)\}$, obey the potentially nonlinear but known dynamics $x(k+1) = f(x(k),u(k))$. Furthermore, they (at least locally) optimally solve the following problem 
\begin{subequations}
\begin{align}
&\min_{u_{i}}\sum_{i = 0}^{N-1}\ell(F_{i}(U,x_{0}), u_{i},L) \label{eq:shortestpathcost}\\
&\qquad g_p(F_{i}(U,x_{0}), u_{i}) \leq 0, \ \ p = 1, \hdots, P \label{eq:shortestpathconstraints}\\
&\qquad x_{0} = x^{*}_d(k) \label{eq:shortestpathinit}\\
&\qquad x_{N} = x^{*}_d(k+N), \label{eq:shortestpathterminalconstraint}
\end{align}
\label{eq:shortestpathopti}%
\end{subequations}
whose objective $\ell(\cdot, \cdot,L)$ includes an unknown, but constant parameter-vector $L$. It can also include $P$ known inequality constraints $g$. The terminal equality constraint~\eqref{eq:shortestpathterminalconstraint}, in accordance with Bellman's Principle of Optimality, allows to formulate the infinite-horizon problem as a shortest path problem of finite length $N$. Additionally, the operator $F_{i}(U,x_{0})$, with $U = [u_{0},\hdots,u_{N-1}]$, is defined as
\begin{equation}
    F_{i}(U,x_{0}) = 
    \begin{dcases}
        x_{0}  & if  \ i = 0, \\[1ex]
        f(F_{i-1}(U,x_{0}),u_{i-1}) & if  \ i \geq 1,
    \end{dcases}
\end{equation}
\noindent which allows for elimination of the associated equality constraint at each time-step, and an expression of the optimization problem in terms of $U$. Introducing the Lagrange multipliers $\lambda_{i} \in \mathbb{R}^{p}$ and $\upsilon \in \mathbb{R}^{n}$, as well as replacing $x_{0}$ with $x^{*}(k)$, the Lagrangian of problem \eqref{eq:shortestpathopti} is given by
\begin{equation}
\begin{split}
    &\mathcal{L}(U,\lambda_{i},\upsilon,L,x^{*}(k),x^{*}(k+N)) =  \upsilon^{\top} \cdot (F_{N}(U,x^{*}(k)) - x^{*}(k+N)) \\& + \sum_{i = 0}^{N-1}\ell(F_{i}(U,x^{*}(k)), u_{i},L) + \lambda_{i}^{\top} \cdot g(F_{i}(U,x^{*}(k)), u_{i}).
\end{split}
\label{eq:lagrangian}
\end{equation}
Finally, to allow for the consideration of potentially sub-optimal data due to noisy observations, the following KKT-based optimization problem 
is solved with respect to the unknown parameter $L$
\begin{equation}
\begin{split}
&\min_{L, \lambda_{d,i},\upsilon_{d}}\sum_{d = 1}^{D}\Vert \nabla_{U}\mathcal{L}(U,\lambda_{d,i},\upsilon_{d},L,x^{*}(k),x^{*}(k+N))\vert_{U = \mathcal{U}_{d}^{*}} \Vert^{2} \\
&\qquad \lambda_{d,i,p} \cdot g_p(F_{i}(\mathcal{U}^{*}_d,x_{d}^{*}(k)), u_{d}^{*}(i)) = 0, \ \ p = 1, \hdots, P, \ d = 1, \hdots, D \\
&\qquad \lambda_{d,i,p} \geq 0, \ \ p = 1, \hdots, P, \ d = 1, \hdots, D. 
\end{split}
\label{eq:shortestpathoptiioc}%
\end{equation}

\section{Problem Description}
\label{sec:problemdescription}
For the remainder of this paper, the available demonstrations are separated into training and validation data. For this purpose, we define $\mathcal{S}_{t}^{*} = \{(\mathcal{X}_{1}^{*},\mathcal{U}_{1}^{*}), \hdots, (\mathcal{X}_{D_{t}}^{*},\mathcal{U}_{D_{t}}^{*})\}$ and 
$\mathcal{S}_{v}^{*} = \{(\mathcal{X}_{D_{t}+1}^{*},$ $\mathcal{U}_{D_{t}+1}^{*}), \hdots, (\mathcal{X}_{D_{t}+D_{v}}^{*},\mathcal{U}_{D_{t}+D_{v}}^{*})\}$ as training and validation data sets respectively, with $D_{d}$ and $D_{v}$ indicating the number of sequences in each set. All of the collected sequences are assumed to be the result of an optimization problem with a time dependent objective, which we express by defining the parameter $L$ in~\eqref{eq:shortestpathopti} as a function depending continuously on time $t \in \mathbb{R}$.
Furthermore, we make the following assumption regarding the structure of the objective: 
\begin{assumption}
It is assumed that the partially unknown objective function $\ell$ is convex for each fixed time instance t, and can be decomposed as
\begin{equation}
    \ell(x_{i}, u_{i},L(t)) = \Theta(t)\cdot\phi(x_{i}, u_{i}),
    \label{eq:costfunctiondecomp}
\end{equation}
where $\Theta(t) = [\theta_{1}(t),\hdots,\theta_{q}(t)] \in \mathbb{R}^{1 \times q}$ describes a $q$ dimensional row vector of unknown continuous, time dependent functions $\theta_{1}(t),\hdots,\theta_{q}(t)$. The column vector $\phi(x_{i}, u_{i}) \in \mathbb{R}^{q}$ is of equal dimension and known. 
\end{assumption}
\begin{remark}
The structure in \eqref{eq:costfunctiondecomp} allows for flexible cost function choices. Convexity of~$\phi(x_i,u_i)$ can ease the formulation of the associated IOC problem, but the assumption is not a strict requirement (see Remark \ref{re:convexityconstraints}). Note that when $\Theta(t)$ is a constant vector and $\phi(x_i,u_i)$ includes squares of states and inputs, the standard quadratic cost function is obtained. 
\end{remark}
The time dependent formulation of the shortest path IOC optimization problem presented in Section~\ref{sec:preliminaries} results in
\begin{subequations}
\begin{align}
&\min_{x_{i}, u_{i}}\sum_{i = 0}^{N-1}\Theta(t)\cdot\phi(F_{i}(U,x_{0}), u_{i}) \label{eq:timedependentshortestpathcost}\\
&\qquad g(F_{i}(U,x_{0}), u_{i}) \leq 0 \label{eq:timedependentshortestpathconstraints}\\
&\qquad x_{0} = x^{*}_d(k) \label{eq:timedependentshortestpathinit}\\
&\qquad x_{N} = x^{*}_d(k+N), \label{eq:timedependentshortestpathterminalconstraint}
\end{align}
\label{eq:timedependentshortestpathopti}%
\end{subequations}
for which we assume to know all involved constraints. Similarly to~\eqref{eq:shortestpathoptiioc}, we can obtain an estimate of $\Theta(t)$ by solving the following optimization problem using the training data in $\mathcal{S}^*_t$
\begin{equation}
\begin{split}
&\min_{\Theta(t), \lambda_{d,i},\upsilon_{d}}\sum_{d = 1}^{D_{t}}\Vert \nabla_{U}\mathcal{L}(U,\lambda_{d,i},\upsilon_{d},\Theta(t),x_{d}^{*}(k),x_{d}^{*}(k+N))\vert_{U = \mathcal{U}_{d}^{*}} \Vert^{2} \\
&\qquad \lambda_{d,i,p} \cdot g_p(F_{i}(\mathcal{U}_{d}^{*},x_{d}^{*}(k)), u_{d}^{*}(i)) = 0, \ \ p = 1, \hdots, P, \ d = 1, \hdots, D_t \\
&\qquad \lambda_{d,i,p} \geq 0, \ \ p = 1, \hdots, P, \ d = 1, \hdots, D_t, 
\end{split}
\label{eq:timedependentshortestpathoptiioc}%
\end{equation}
with its Lagrangian defined as 
\begin{equation}
\begin{split}
    &\mathcal{L}(U,\lambda_{d,i},\upsilon_{d},\Theta(t),x_{d}^{*}(k),x_{d}^{*}(k+N)) = \upsilon_{d}^{\top} \cdot (F_{N}(U,x_{d}^{*}(k)) - x_{d}^{*}(k+N)) + \\& \sum_{i = 0}^{N-1}\Theta(t)\cdot\phi(F_{i}(U,x_{d}^{*}(k)), u_{i}) + \lambda_{d,i}^{\top} \cdot g(F_{i}(U,x_{d}^{*}(k)), u_{i}).
\end{split}
\label{eq:timedependentlagrangian}
\end{equation}
Optimizing over a vector of unknown continuous, time dependent functions $\theta_{1}(t),\hdots,\theta_{q}(t)$ results in an intractable problem. For this reason, in the following section, we define a model for $\Theta(t)$ consisting of a linear combination of trigonometric basis functions, so that problem~\eqref{eq:timedependentshortestpathoptiioc} can be reformulated as an optimization with respect to a finite set of hyperparameters, i.e. the sinusoids' frequencies. 
This allows for overcoming the intractability of estimating a (potentially very long) sequence of time-varying parameters and provides an efficient framework for identifying the relation between cost and time. \\
The quality of the estimated parameter $\hat{\Theta}(t)$ is evaluated by solving the forward optimization control problem  fixing $\hat{\Theta}(t)$, for each initial state $x_{d}^{*}(k), \; d \in [D_{t}+1, D_{t}+D_{v}]$, in the validation set $\mathcal{S}^*_v$. The optimized sequences are indicated with $\hat{\mathcal{X}}_{d} = \{\hat{x}_{d}(k), \hdots, \hat{x}_{d}(k+N)\}$ and $\hat{\mathcal{U}}_{d}= \{\hat{u}_{d}(k), \hdots, \hat{u}_{d}(k+N-1)\}$, and collected in $\hat{\mathcal{S}}_{v} = \{(\hat{\mathcal{X}}_{D_{t}+1},\hat{\mathcal{U}}_{D_{t}+1}), \hdots, (\hat{\mathcal{X}}_{D_{t}+D_{v}},\hat{\mathcal{U}}_{D_{t}+D_{v}})\}$. Consequently, the validation error is defined as an averaged root mean square error between the optimized sequences in $\hat{\mathcal{S}}_v$ and the original validation sequences in $\mathcal{S}^*_v$
\begin{equation}
\begin{split}
    e_{v}(\hat{\mathcal{S}}_{v},\mathcal{S}_{v}^{*}) = \frac{1}{D_{v}} \sum_{d=D_{t} + 1}^{D_{t}+D_{v}}(\frac{1}{N}\sum_{k=0}^{N-1}
    \Vert\hat{x}_{d}(k) - x^{*}_{d}(k)\Vert^{2}+\Vert\hat{u}_{d}(k) - u^{*}_{d}(k)\Vert^{2})^{\frac{1}{2}}.
    \label{eq:validationerror}
\end{split}
\end{equation}

\subsection*{The Intuition of Using a ``Sliding-window" Approach and Where it Fails}
\label{subsec:filterbased}
\begin{wrapfigure}{r}{0.28\linewidth}
\centering
\vspace{-1.0em}
\includegraphics[scale=0.3]{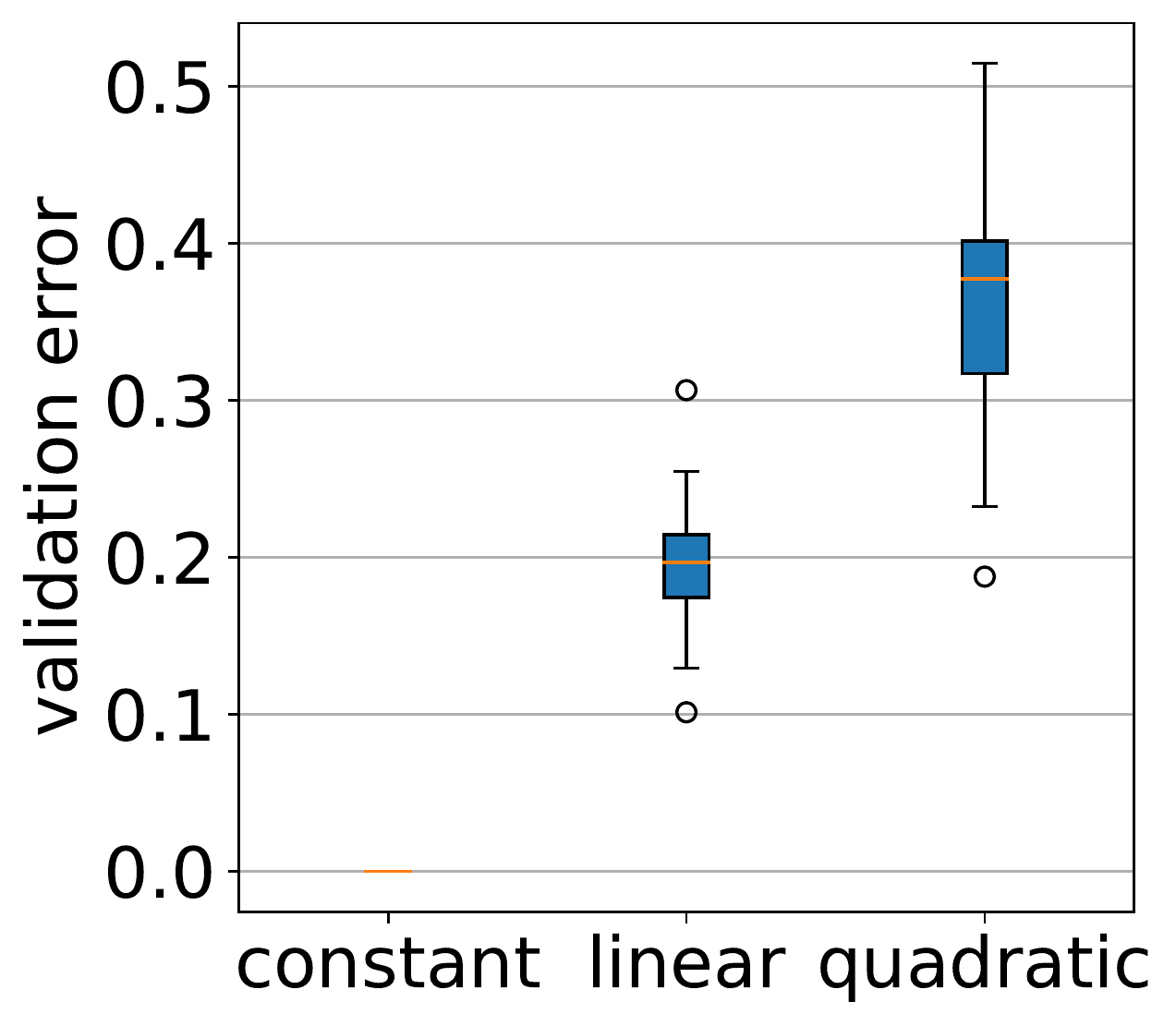}
\vspace{-1.0em}
\caption{the effect of increasing nonlinearity of $\Theta(t)$}
\vspace{-2.4em}
\label{fig:filtererror}
\end{wrapfigure}

In the previous section, we provided a raw formulation of the time dependent spIOC in~\eqref{eq:timedependentshortestpathoptiioc}, and discussed its intractability when optimized as a batch problem. A naive approach for overcoming this consists in solving~\eqref{eq:timedependentshortestpathoptiioc} sequentially, i.e. using a linear Kalman filter to estimate the time dependent parameters $\Theta(t)$, exploiting the fact that the result in~\cite{menner2019} allows for partitioning the demonstrator sequences in small windows. The idea is to use a ``sliding window" approach, in which we iteratively re-estimate a portion $\theta_M \in \mathbb{R}^{1 \times M}$ of the vector $ \Theta(t) $. We assume that the estimation window $\theta_M$ stays constant over the window length $M$, and choose its length in order to keep the estimation problem well-defined. The following example aims at offering an intuitive explanation of when this approach fails, therefore motivating the choice of learning a model for time dependency. In particular, we consider a one-element spring-damper system.
As shown in Figure \ref{fig:filtererror}, as the order of time-dependency in $\Theta(t)$ increases, the validation error computed as in~\eqref{eq:validationerror} grows due to the inability of the sliding window approach to capture the increasing degree of nonlinearity in time.

\section{Trigonometric Time Dependent IOC}
\label{sec:tblrbasedtimedependentioc}
In the following section we present a strategy to address the limitations of a recursive parameter estimation discussed in  Section~\ref{sec:problemdescription}, by explicitly introducing a model for time dependency. The idea is to construct a model using a linear combination of trigonometric basis functions and to optimize their hyperparameters within an IOC framework. By learning a model of the cost function's relation with time, we can exploit its predictive capabilities for evaluation on unseen time instances. We refer to this approach as
trigonometric time dependent IOC approach (TTD-IOC). 
\subsection{Trigonometric Basis Functions as Time Features}
\label{sec:tblrintroduction}
The model we consider for each time dependent cost function parameter is
\begin{equation}
\begin{split}
    \theta_{m}(t) = &\alpha_{1,m}1 + \alpha_{2,m}\cos{(\omega_{1}t)} + \alpha_{3,m} \sin{(\omega_{1}t)} + \\& \hdots + \alpha_{2E,m}\cos{(\omega_{E}t)} + \alpha_{2E+1,m} \sin{(\omega_{E}t)},
\end{split}
\label{eq:tfeaturedependentcostfunctionparametermodel}
\end{equation}
\noindent consisting of $2E$ basis functions, defined by the set of frequencies $\mathcal{W} = \{\omega_{1}, \hdots, \omega_{E}\}$. Potential offsets are modelled via an additional constant bias.
We define the basis functions vector $\Omega(\mathcal{W},t) \in \mathbb{R}^{1\times (2E+1)}$ as a row vector consisting of $2E+1$ elements
\begin{equation}
\begin{split}
    \Omega(\mathcal{W},t) = [1, \cos{(\omega_{1}t)}, \sin{(\omega_{1}t)}, \hdots, \cos{(\omega_{E}t)}, \sin{(\omega_{E}t)}],
\end{split}
\end{equation}

\noindent together with the matrix $A \in \mathbb{R}^{(2E+1) \times q}$ consisting of parameters that linearly combine the basis functions
\begin{equation}
  A_{(2E+1)\times q} =
  \left[ {\begin{array}{ccc}
    \alpha_{11} & \hdots & \alpha_{1q} \\
    \vdots & \ddots & \vdots \\
    \alpha_{(2E+1)1} & \hdots & \alpha_{(2E+1)q} \\
  \end{array} } \right],
\end{equation}

\noindent resulting in the following model for the time dependent vector $\Theta(t)$ as
\begin{equation}
    \Theta(t) = \Omega(\mathcal{W},t)A.
    \label{eq:tblrfeaturedecomp}
\end{equation}
\vspace{-1.5em}
\begin{remark}
    The model in \eqref{eq:tfeaturedependentcostfunctionparametermodel} assumes that all cost parameter time dependencies can be approximated with the same set of frequencies $\mathcal{W}$. Considering individual frequencies for each time dependent cost parameter is possible, however, it increases the number of unknown parameters and, accordingly, the amount of required data. 
\end{remark}

\subsection{Regularized Optimization Problem for TTD-IOC}
\label{sec:featuredependentioc}
In the following, we present the proposed algorithm for optimizing the hyperparameters $\mathcal{W}$ and $A$ and introduce the regularization term for obtaining a sparse solution. 
\noindent For this purpose we include the model \eqref{eq:tblrfeaturedecomp} into the optimization problem presented in equation \eqref{eq:timedependentshortestpathoptiioc} and consider that measurements are only available for time instances $t=kT_s$, where $T_s$ indicates the sampling time. Additionally, the objective is extended with a lasso inspired regularizer $\beta \in \mathbb{R}_{+}$. Defining $\mathcal{L}_{U,d} =  \mathcal{L}(U,\lambda_{d,i},\upsilon_{d},\Omega (\mathcal{W},kT_s)A,x^{*}_{d}(k),x^{*}_{d}(k+N))$, the regularized, time dependent Lagrangian optimization problem results in 
\begin{equation}
\begin{split}
\hat{\mathcal{W}}, \hat{A}&=\text{arg}\min_{\mathcal{W}, A, \lambda_{d,i},\upsilon_{d}}\sum_{d = 1}^{D_{t}}\Vert (\nabla_{U}\mathcal{L}_{U,d})\vert_{U = \mathcal{U}_{d}^{*}} \Vert^{2} + \beta\vert A(2:2E+1,2:q) \vert \\
&\qquad \lambda_{d,i,p} \cdot g_p(F_{i}(\mathcal{U}_{d}^{*},x_{d}^{*}(k)), u_{d}^{*}(i)) = 0, \ \ p = 1, \hdots, P, \ d = 1, \hdots, D_t \\
&\qquad \lambda_{d,i,p} \geq 0, \ \ p = 1, \hdots, P, \ d = 1, \hdots, D_t \\
&\qquad A(:,0) = v_{\alpha}^{*}.
\end{split}
\label{eq:tblrshortestpathoptiregularized}%
\end{equation}
\begin{wrapfigure}{L}{0.35\textwidth}
    \begin{minipage}{0.35\textwidth}
        \begin{algorithm}[H]
            \caption{TTD-IOC}
            \label{alg:featuredependetioc}
            $\text{Define} \ \beta_{i}, \beta_{f},\beta_{s},D_{t}, D_{v},\bar{e}_{\beta}$\\
                $\beta = \beta_{i}$ \\
            \While{$\beta \leq \beta_{f}$}
                {
                $\hat{\mathcal{W}}_{\beta}, \hat{A}_{\beta} \gets solve(\ref{eq:tblrshortestpathoptiregularized}) \; \text{with } \beta$\\
                $\hat{\Theta}_{\beta}(t) \gets \Omega(\hat{\mathcal{W}}_{\beta},t)\hat{A}_{\beta}$ \\
                $\hat{\mathcal{S}}_{v} \gets solve(\ref{eq:timedependentshortestpathopti}) \ w.r.t. \ \hat{\Theta}_{\beta}(t)$\\
                $e_{\beta} \gets solve(\ref{eq:validationerror})$\\
                \If{$e_{\beta} \leq \bar{e}_{\beta}$}
                    {$\hat{\mathcal{W}}, \hat{A} = \hat{\mathcal{W}}_{\beta}, \hat{A}_{\beta}$ \\
                    $\bar{e}_{\beta} = e_{\beta}$}
                $\beta=\beta+\beta_{s}$}
            $\hat{\Theta}(t) \gets \Omega(\hat{\mathcal{W}},t)\hat{A}$
        \end{algorithm}
    \end{minipage}
    \vspace{-1.5em}
\end{wrapfigure}

\noindent The lasso-inspired regularizer thereby penalizes the sum of the absolute values of the sub-matrix of $A$ to minimize the amount of time-varying cost parameters, offering an automatic selection of the model complexity. 
Furthermore, the trivial solution of setting all elements of $A$ equal to zero is excluded from the set of potential solutions by adding the equality constraint $A(:,1) = v_{\alpha}^{*}$, where $v_{\alpha}^{*} \in \mathbb{R}^{2E+1}$ is user-defined. Finally, the estimated feature dependent cost parameter are obtained as $\hat{\Theta}(t) = \Omega(\hat{\mathcal{W}},t)\hat{A}$. Note that the mentioned predictive capabilities with respect to unseen time instances allow for considering different sampling time instances for further predictions.
\begin{remark}
Depending on the cost features $\phi(x_{i}, u_{i})$, further constraints can be added to the proposed optimization problem in \eqref{eq:tblrshortestpathoptiregularized} to preserve the convexity of the estimated cost function $\ell(\cdot, \cdot, \cdot)$ for each fixed time instance t, e.g. non-negativity constraints on $\Theta(t)$ for convex $\phi(x_{i}, u_{i})$. 
\label{re:convexityconstraints}
\end{remark}
The proposed algorithm, presented in Algorithm \ref{alg:featuredependetioc}, consists of a line search over $\beta$, between $\beta_{i}$ and $\beta_{f}$, with a step-wise increase of $\beta_{s}$. For all values of $\beta$ the optimization problem in \eqref{eq:tblrshortestpathoptiregularized} is solved, and the quality of the respective parameter estimate is evaluated with respect to the resulting validation error introduced in \eqref{eq:validationerror}. The initial value of $\bar{e}_{\beta}$ is chosen sufficiently high, making sure that it is adjusted in the algorithm's first iteration.

\section{Results}
\label{sec:results}
\begin{figure}[htb]
    \centering
    \begin{minipage}[t]{0.45\linewidth}
        \centering
        \vspace{-4.3em}
        \includegraphics[width=.95\linewidth]{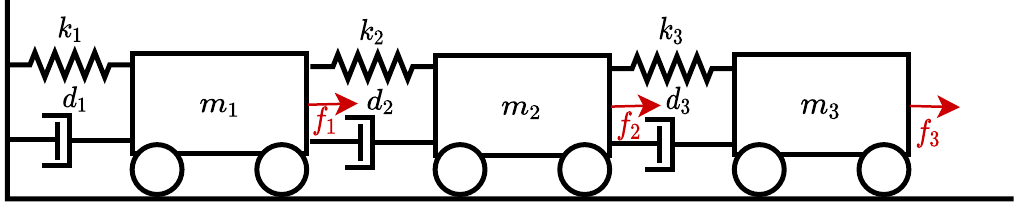}
    \end{minipage}
    \begin{minipage}[t]{0.45\linewidth}
        \centering
        \includegraphics[width=.65\linewidth]{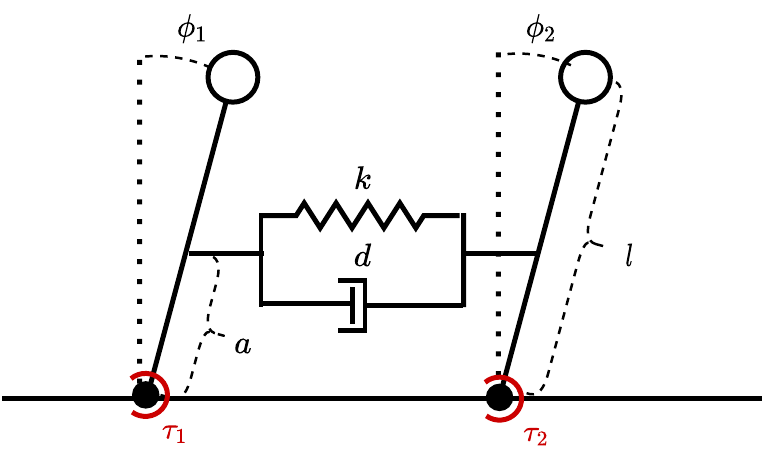}
    \end{minipage}
    \vspace{-1.0em}
    \caption{Illustration of considered dynamical systems. The multi-layer spring-damper system is depicted on the left and the double pendulum on the right.}
    \label{fig:simulationsystem}
    \vspace{-1.6em}
\end{figure}

In this section, we analyse the proposed algorithm by applying it to two illustrative simulation scenarios: the first is a linear dynamical system, i.e. a three-layer spring-damper system (sys1)
, while the second system consists of two inverted pendulums (sys2) connected via a spring-damper element, as an example of nonlinear dynamics (see Figure~\ref{fig:simulationsystem}).
\vspace{-0.7em}
\noindent For both systems, we choose the cost function parameter associated with one input to vary with time and design $\phi(x,u)$ to include squares of all states and inputs. 
The cost parameters of the considered forward problems from which we obtain training and validation demonstrations are presented in Table \ref{tab:parameters}. For the time dependent cost parameter $\theta_{m,i}$ we define three different continuous functions
\begin{subequations}
\begin{align*}
    & \theta_{m,1}(t)= 4 + 1.5\cos{(2t)}+1.5\cos{(3t)} \\
    & \theta_{m,2}(t)= 1.5+0.02t^{2}-0.01t \\
    & \theta_{m,3}(t)= 4 + e^{0.2t}
\end{align*}
\end{subequations}
Results are obtained using a HP ProBook 440 with an Intel Core i7 processor, while using the Ipopt optimization framework \cite{wachter2006} within the Casadi framework~\cite{Andersson2019}. The issue of dealing with a nonconvex optimization problem is thereby addressed via a grid search over suitable frequency values ($\omega_{init}$) for the initialization of the applied solver. Its step size and the grid corners can be adjusted by altering the values of $\omega_{i}$, $\omega_{f}$,  and $\omega_{s}$. For the subsequent experiments with sys1 they are chosen as $0.5,2.5$, and $2.0$ and as $0.11,0.11$, and $0.0$ for sys2, respectively. The line search parameters are set to $\beta_i = 0.04,\beta_f = 0.06$, and $\beta_s=0.01$ for sys1 and to $\beta_i = 0.058,\beta_f = 0.061$, and $\beta_s=0.003$ for sys2. 
The performance of our proposed approach is evaluated on each system with respect to the validation error defined in~\eqref{eq:validationerror}.

\begin{small}
    \begin{table}[h]
        \begin{center}
        \begin{tabular}{ | c | c | c | c | c | c | c | c | c | c |}
        \hline
          & $\theta_{x_1/\phi_1}$ & $\theta_{\dot{x}_1/\dot{\phi}_1}$ & $\theta_{x_2/\phi_2}$ & $\theta_{\dot{x}_2/\dot{\phi}_2}$ & $\theta_{x_3}$ & $\theta_{\dot{x}_3}$ & $\theta_{f_1/\tau_1}$ & $\theta_{f_2/\tau_2}$ & $\theta_{f_3}$  \\ \hline \hline
        $\bold{sys1}$ & 7 & 5 & 6 & 8 & 6.5 & 5.5 & 2 & 4 & $\theta_{m,i}$  \\ \hline
        $\bold{sys2}$ & 7 & 5 & 10 & 5 & - & - & 4 &$ \theta_{m,i}$ & - \\ \hline
        \end{tabular}
        \caption{
        Cost parameters of considered forward problems.}
        \vspace{-2.5em}
        \label{tab:parameters}
        \end{center}
    \end{table}
\end{small}

\subsection{Multi-layer Spring-Damper System}
\label{subsec:multilayerspringdamper}
The considered multi-layer spring-damper system consists of three stacked mass elements, $m_1$, $m_2$, and $m_3$, connected to each other or the wall by a spring-damper pair with spring constants $k_1$ up to $k_3$ and damping constants $d_1$ up to $d_3$. The input is given by a force vector $F = [f_1,f_2,f_3]$ consisting of three forces that can be exerted on their respective mass. An illustration of the system is given in the left part of Fig. \ref{fig:simulationsystem} and the dynamics by equation \eqref{eq:multilayerspringdampersystemdyn}. 
\begin{equation}
\begin{split}
    &\ddot{x}_{1} = {m_1}^{-1}(-(k_{1}+k_{2}){x}_{1} -(d_{1}+d_{2})\dot{x}_{1} + k_{2}{x}_{2} + d_{2}\dot{x}_{2} + f_{1})  \\
    &\ddot{x}_{2} = {m_2}^{-1}(k_{1}{x}_{1}+ d_{1}\dot{x}_{1}-(k_{2}+k_{3}){x}_{2}-(d_{2}+d_{3})\dot{x}_{2} + k_{3}{x}_{3} + d_{3}\dot{x}_{3} + f_{2})\\
    &\ddot{x}_{3}  = {m_3}^{-1}(k_{3}{x}_{2} +  d_{3}\dot{x}_{2} -k_{3}{x}_{3} -d_{3}\dot{x}_{3} + f_{3})\
\end{split}
\label{eq:multilayerspringdampersystemdyn}
\end{equation}
The proposed modelling of time dependency within the presented approach enables an improvement with respect to the vanilla spIOC described in Section~\ref{sec:preliminaries}. We measure the improvement in terms of validation error, and confirm the ability of the learned cost to generalize for unseen scenarios, i.e. different initial conditions, as can be observed in the left plots of Figure~\ref{fig:validationerrorssys1and2}.

\subsection{Inverted Double Pendulum}
\label{subsec:inverteddoublependulum}
\begin{figure}[!h]
    \begin{minipage}[t]{0.24\linewidth}
    \includegraphics[width=.99\linewidth]{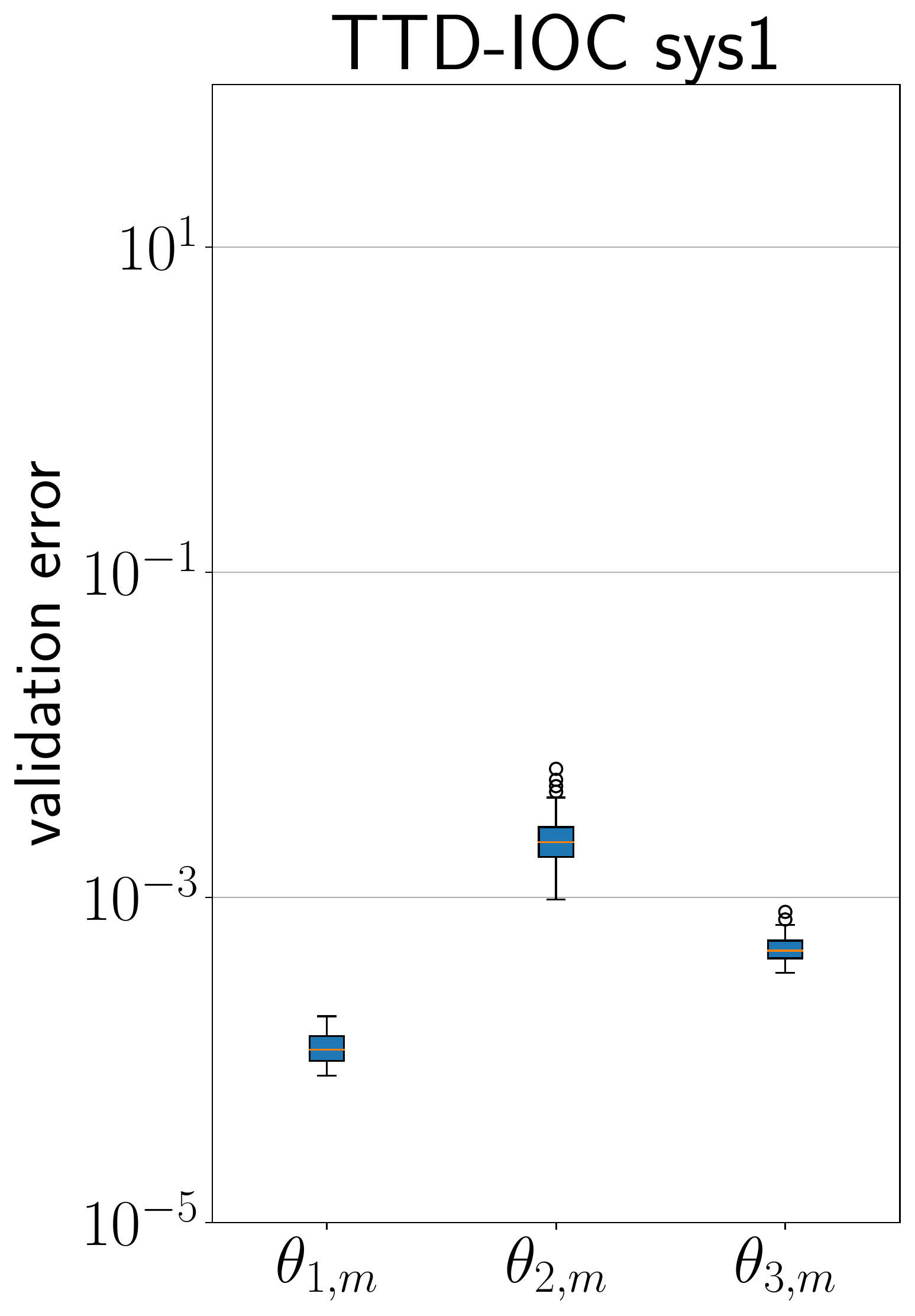}
    \end{minipage}
    \begin{minipage}[t]{0.24\linewidth}
    \includegraphics[width=.99\linewidth]{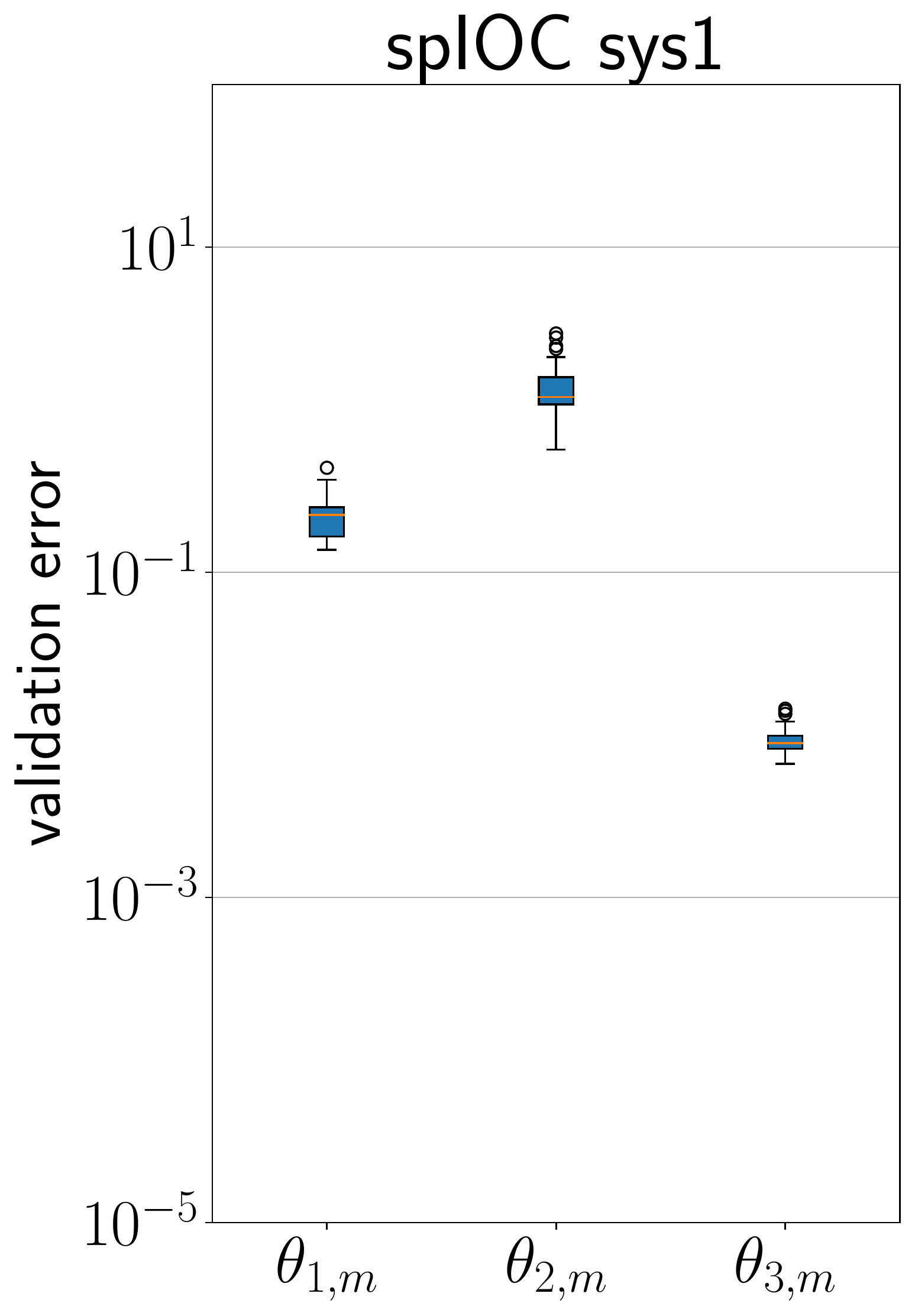}
    \end{minipage}
    \begin{minipage}[t]{0.24\linewidth}
    \includegraphics[width=.99\linewidth]{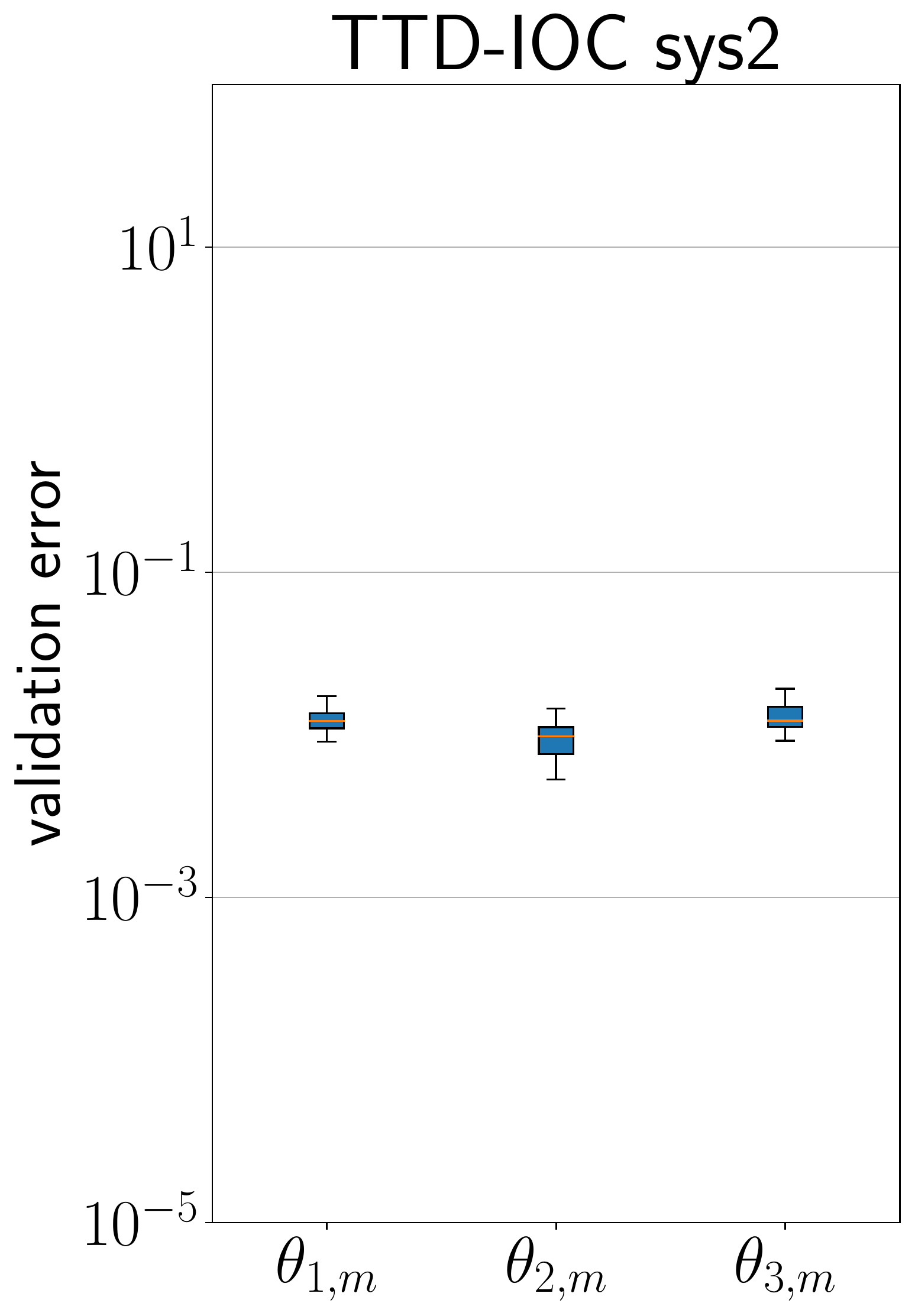}
    \end{minipage}
    \begin{minipage}[t]{0.24\linewidth}
    \includegraphics[width=.99\linewidth]{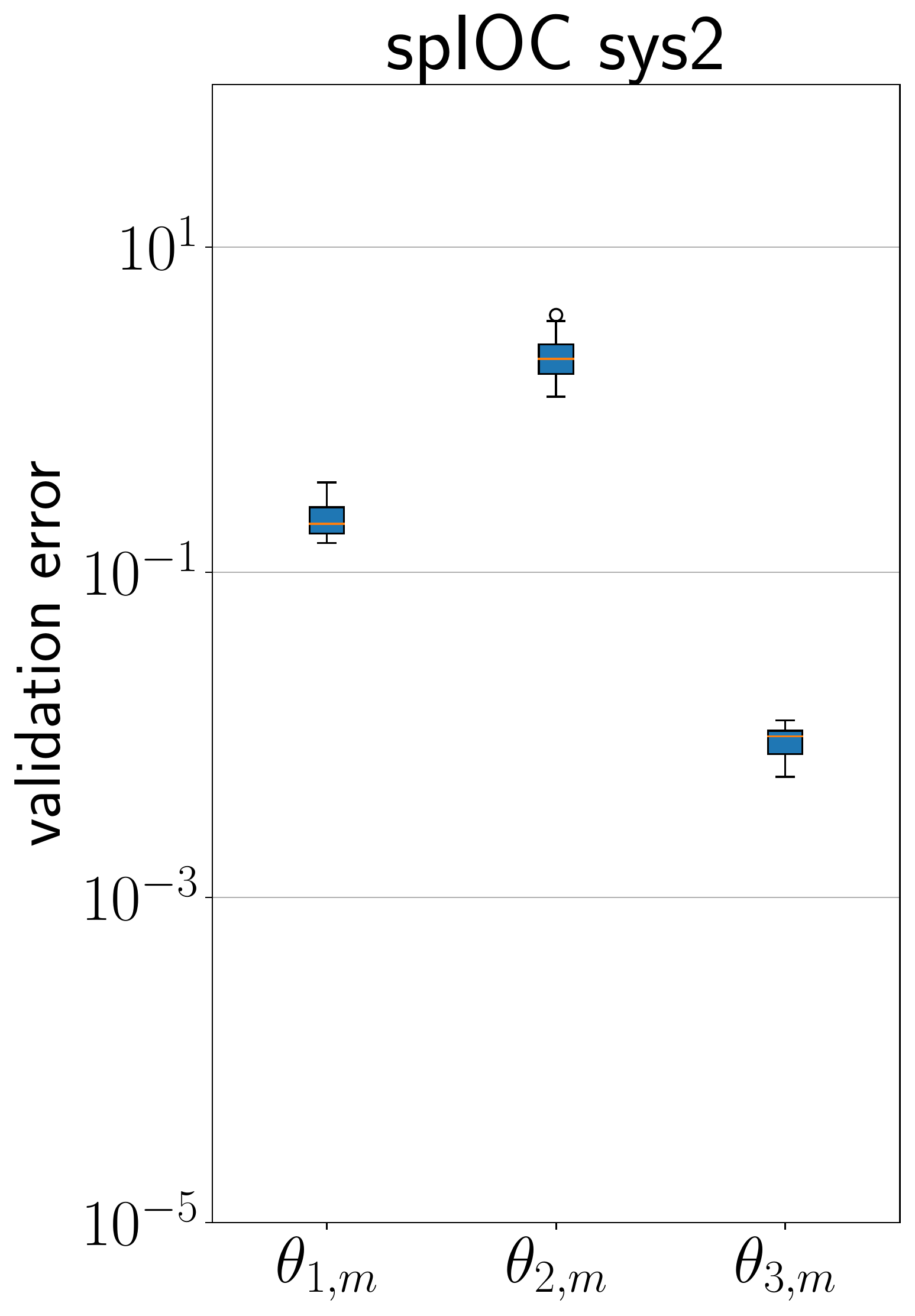}
    \end{minipage}
    \vspace{-1.0em}
    \caption{Comparison between the validation errors produced by the trigonometric time dependent IOC (TTD-IOC) and the shortest path IOC (spIOC) for time dependent objectives for both sys1 and sys2. Results are shown on a log scale. }
    \label{fig:validationerrorssys1and2}
\end{figure}

The inverted double pendulum consists of two inverted pendulums of length $l$, having a mass element $m$ attached at its top. At the height indicated with $a$ they are connected via a spring-damper pair whose constants are indicated by $k$ and $d$. Each pendulum can be actuated by its individual torque $\tau_{1}$ or $\tau_{2}$. An illustration of the system is given in the right picture of Fig. \ref{fig:simulationsystem} and the dynamics by equation \eqref{eq:doublependulumdyn}. Defining $F = ka(\sin{(\phi_2)}-\sin{(\phi_1)}) + da(\cos{(\phi_2)}\dot{\phi_2} - \cos{(\phi_1)}\dot{\phi_1})$ it follows
\begin{equation}
\begin{split}
    &\ddot{\phi}_{1} = g{l}^{-1}\sin{(\phi_1)} +a\cos{(\phi_1)}F+(ml^{2})^{-1}\tau_{1}  \\
    &\ddot{\phi}_{2} = g{l}^{-1}\sin{(\phi_2)} -a\cos{(\phi_2)}F+(ml^{2})^{-1}\tau_{2}.
\end{split}
\label{eq:doublependulumdyn}
\end{equation}
The resulting validation errors for our considered nonlinear system are given in the right plots of Figure \ref{fig:validationerrorssys1and2}. 
Similarly to the previous example, the proposed approach shows again improved validation errors with respect to spIOC. 

\subsection{Comparative Study}

\begin{wrapfigure}{r}{0.35\linewidth}
\centering
\vspace{-1.5em}
\includegraphics[scale=0.27]{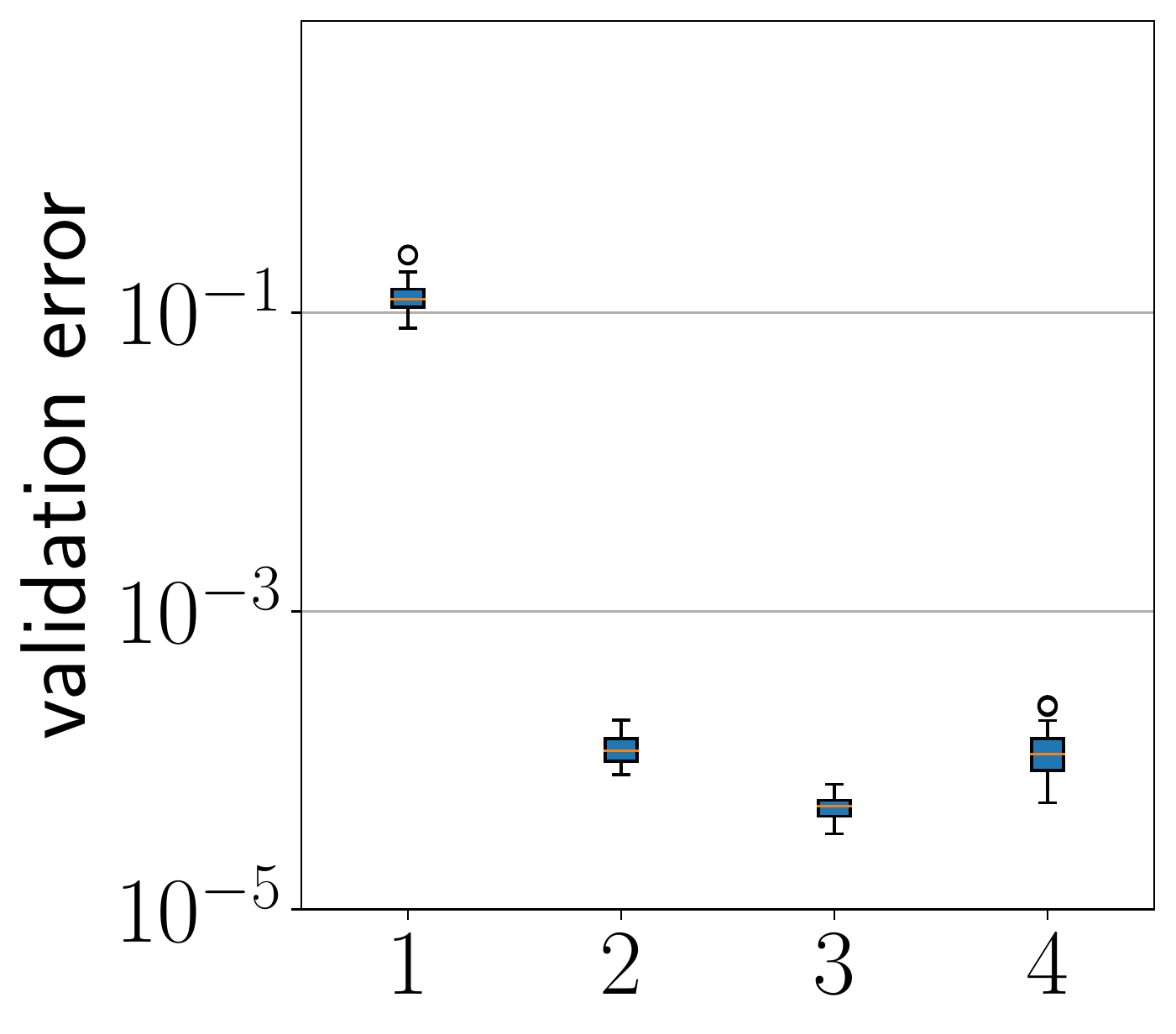}
\vspace{-1.5em}
\caption{The effect of varying $E$.}
\vspace{-1.1em}
\label{fig:featureerror}
\end{wrapfigure}

\begin{figure}[htb]
    \begin{minipage}[t]{0.33\linewidth}
    \includegraphics[width=.79\linewidth]{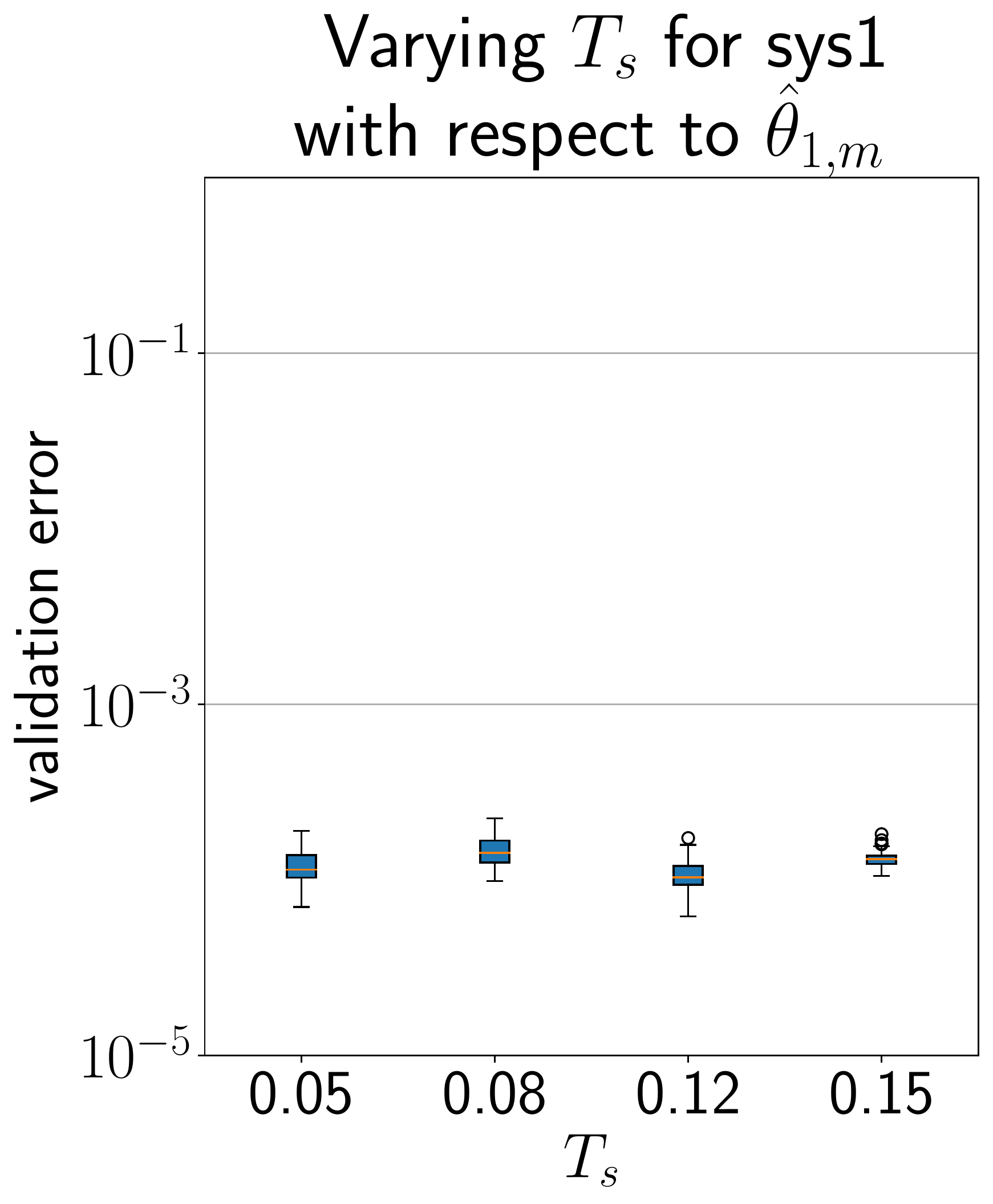}
    \end{minipage}
    \begin{minipage}[t]{0.33\linewidth}
    \includegraphics[width=.79\linewidth]{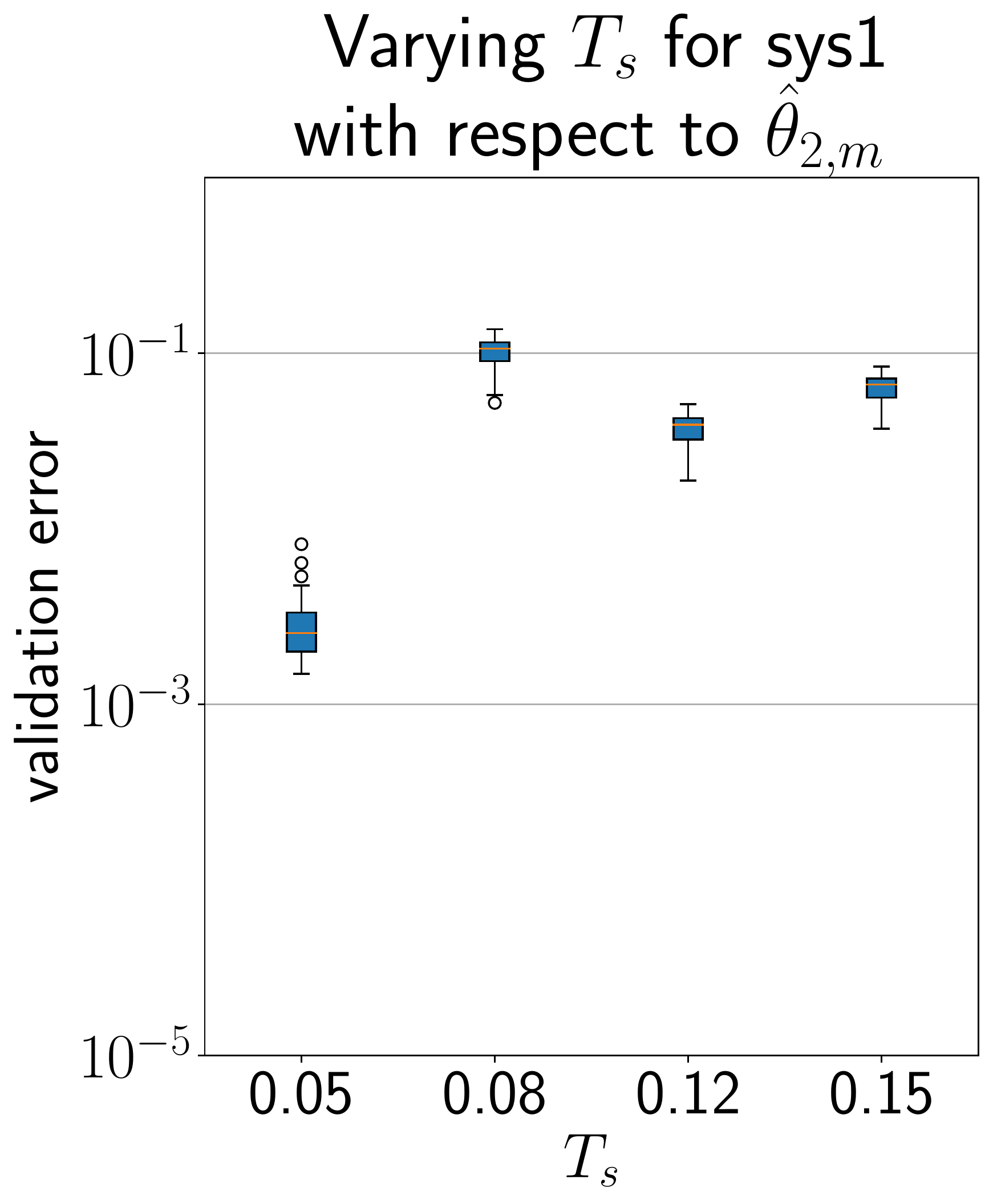}
    \end{minipage}
    \begin{minipage}[t]{0.33\linewidth}
    \includegraphics[width=.79\linewidth]{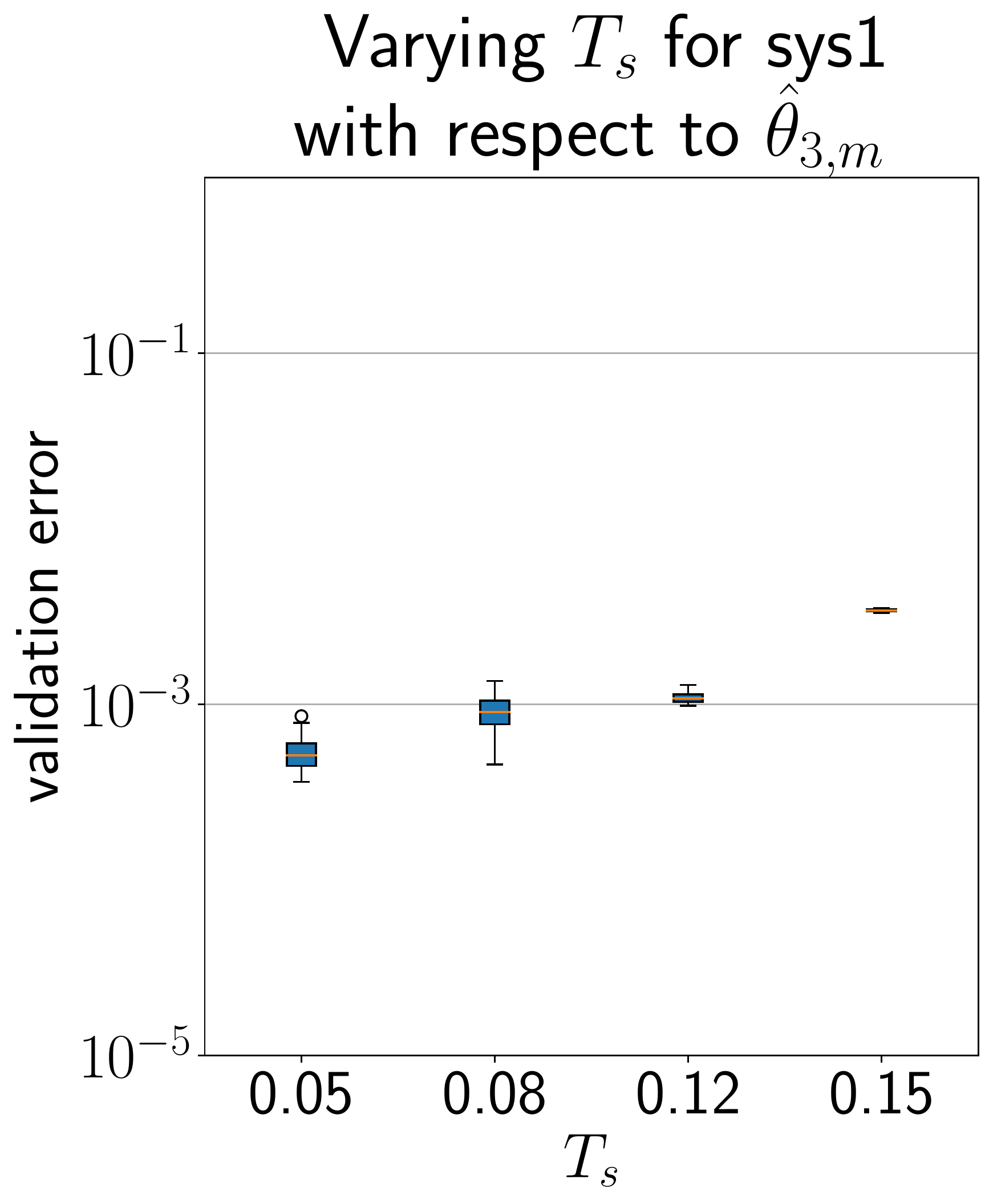}
    \end{minipage}
    \vspace{-2.0em}
    \caption{Validation error of using $\hat{\theta}_{1,m}$ up to $\hat{\theta}_{3,m}$ for optimal sequences of sys1 that vary in $T_s$.} 
    \label{fig:validationerroranalysists}
\end{figure}

\begin{figure}[htb]
    \begin{minipage}[t]{0.33\linewidth}
    \includegraphics[width=.79\linewidth]{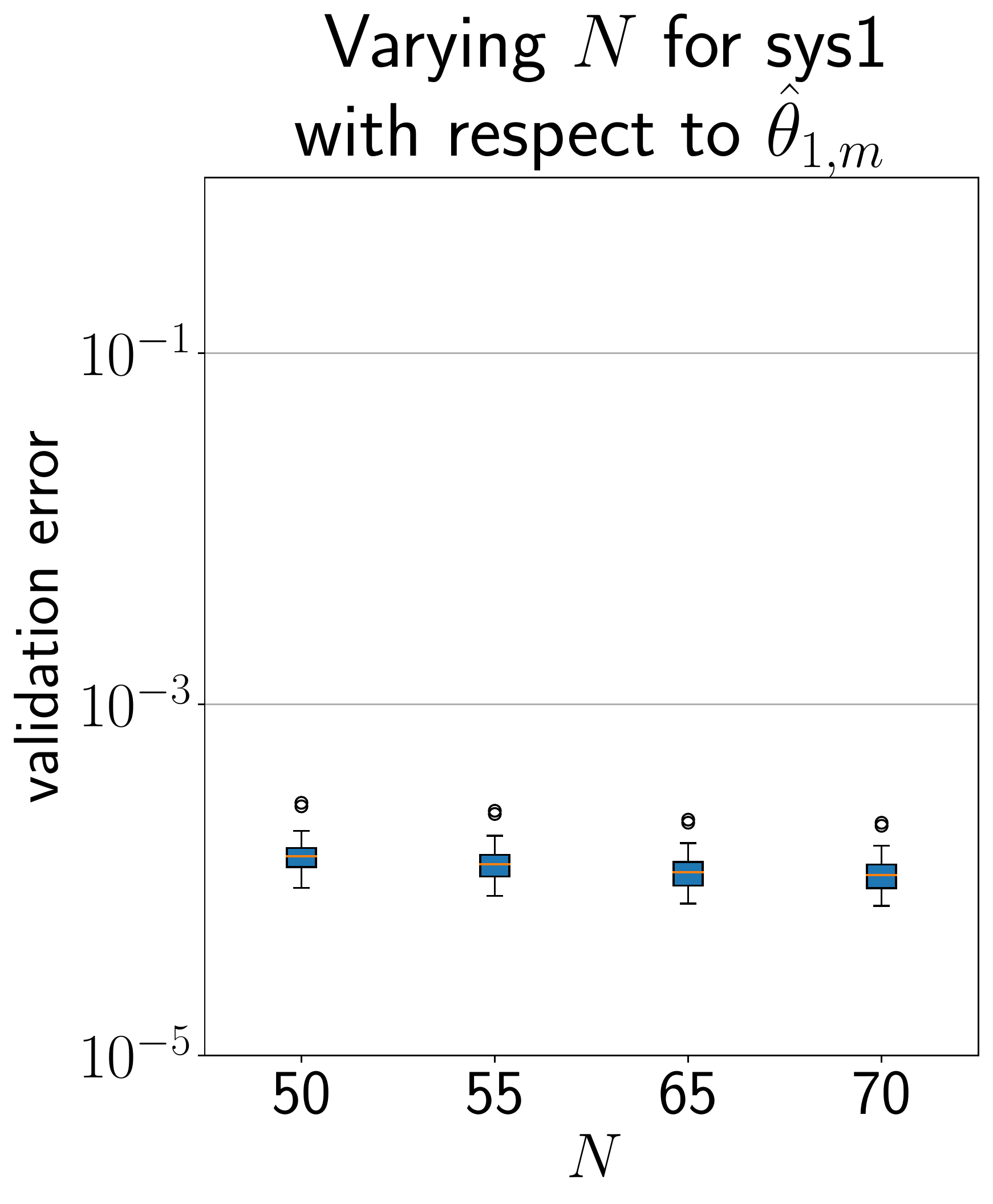}
    \end{minipage}
    \begin{minipage}[t]{0.33\linewidth}
    \includegraphics[width=.79\linewidth]{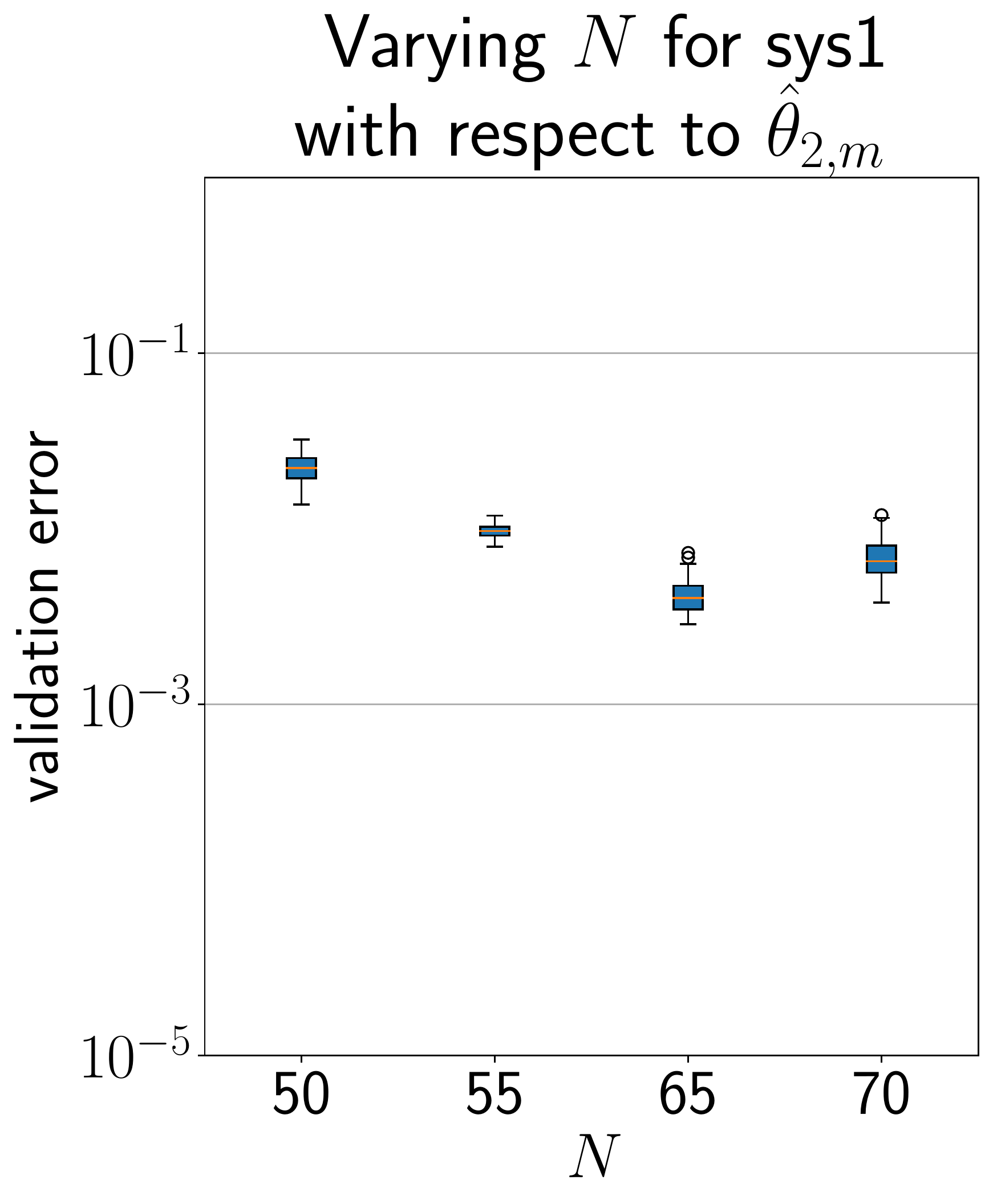}
    \end{minipage}
    \begin{minipage}[t]{0.33\linewidth}
    \includegraphics[width=.79\linewidth]{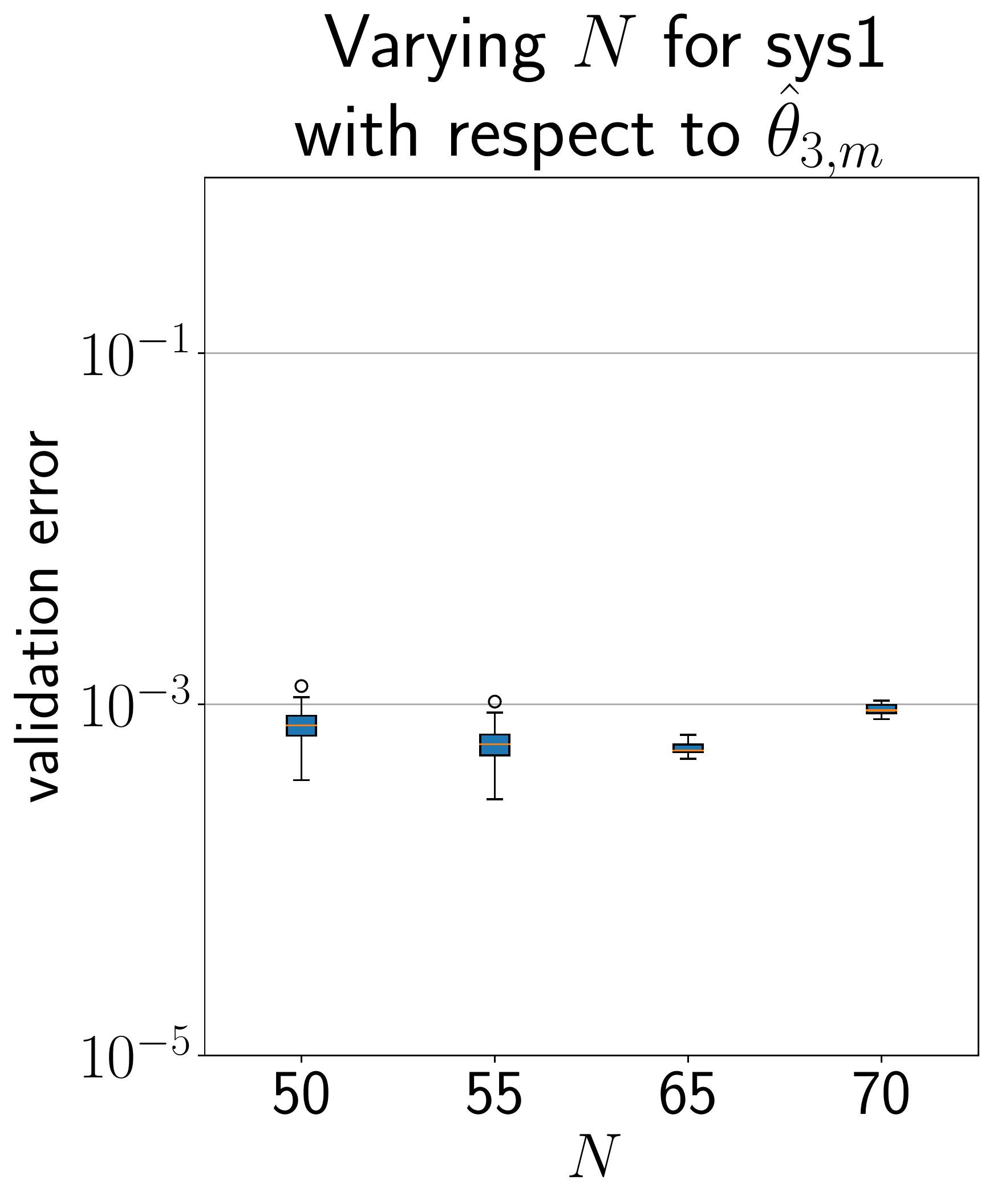}
    \end{minipage}
    \vspace{-2.0em}
    \caption{Validation error of using $\hat{\theta}_{1,m}$ up to $\hat{\theta}_{3,m}$  for optimal sequences of sys1 that vary in $N$.}
    \label{fig:validationerroranalysisN}
    \vspace{-1.0em}
\end{figure}

In this section, we investigate the effects of varying sampling times and horizon lengths on the validation error. For this purpose, the parameter estimates of sys1, obtained in accordance with training data sequences $\mathcal{S}_{t}^{*}$ with a horizon length of $N=60$ and sampling time $T_s = 0.1$, are used to obtain optimal sequences for an adjusted horizon length or an adjusted sampling time. They are evaluated with respect to new and unseen validation data of given $N$ and $T_s$. The validation errors resulting from a variation in sampling time are depicted in Figure \ref{fig:validationerroranalysists}, and those for different horizon lengths in Figure~\ref{fig:validationerroranalysisN}. Furthermore, we tested the performance of the approach in the presence of model mismatch, specifically when the number of chosen basis functions E is different from the true model. Assuming that the true cost function is described by $2$ trigonometric features, the value $E$ is varied in the set $\{1,2,3,4\}$, and the obtained validation errors are displayed in Figure~\ref{fig:featureerror}.


\subsection{Discussion}
\label{subsec:discussion}
The results obtained in Section \ref{subsec:multilayerspringdamper} and \ref{subsec:inverteddoublependulum} show an increased ability to mimic the optimal input and output sequences of underlying continuously time dependent objective functions in comparison to an existing IOC approach that does not model time dependent features (spIOC). Furthermore, in Figure~\ref{fig:validationerroranalysists} and  Figure~\ref{fig:validationerroranalysisN}, it can be observed that modeling the time dependency as an explicit feature allows for the consideration of different sampling times and horizon lengths.
In general, a better estimate in terms of validation error results in better generalization capabilities. The investigation of varying the number of frequencies in the considered trigonometric feature vector indicates that, given the regularization parameter $\beta$, a larger value of $E$ does not lead to a degradation in terms of the validation error. Furthermore, by inspecting the elements of $A$ it can be seen that the parameters associated with unused basis functions are close to zero. While the validation error increases when choosing fewer frequencies, the proposed approach still outperforms the spIOC estimate. 
\section{Conclusion}
\label{sec:conclusion}
This paper has presented a procedure for modeling time dependency as an explicit cost function feature using linear combinations of trigonometric basis functions. Building on previous results, we extended the shortest path IOC approach to include the additional optimization over the time feature hyperparameters, and discussed its performance by analysing two simulation examples. Results show lower validation error compared with the shortest path IOC approach. Furthermore, the use of a model provides an estimate for not only seen but also unseen instances in time, and the proposed method thereby also provides high-quality predictions (i.e. low validation errors) when varying the sampling time, as well as the horizon length in the forward optimization problem. 
In the future, we plan to examine further the effect of the amount of training data on the validation error and to compare the proposed solution strategy for addressing the nonconvexity of the Lagrangian optimization problem against, e.g., sampling-based routines, as well as test the proposed procedure in real-world problems.

\newpage


\bibliography{ioc_feature_dependence}

\end{document}